\documentclass[twocolumn,amsmath,amssymb,prl,floatfix]{revtex4}
\usepackage{graphicx}
\usepackage{dcolumn}
\usepackage{bm}
\begin{document}

\title{Probing fractal magnetic domains on multiple length scales in Nd$_{2}$Fe$_{14}$B}

\author{A.~Kreyssig$^{1,2,3}$}
\email{kreyssig@ameslab.gov}
\author{R.~Prozorov$^{1,2}$}
\author{C.~D.~Dewhurst$^{4}$}
\author{P.~C.~Canfield$^{1,2}$}
\author{R.~W.~McCallum$^{1}$}
\author{A.~I.~Goldman$^{1,2}$}

\affiliation{\\$^1$Ames Laboratory, US DOE; Ames, IA 50011; USA}

\affiliation{$^2$Department of Physics and Astronomy; Iowa State
University; Ames, IA 50011; USA}

\affiliation{$^3$Institut f\"ur Festk\"orperphysik; Technische
Universit\"at Dresden; 01062 Dresden; Germany}

\affiliation{$^4$Institut Laue-Langevin; 6, rue Jules Horowitz;
38042 Grenoble; France}


\begin{abstract}
Using small-angle neutron scattering, we demonstrate that the 
complex magnetic domain patterns at the surface of 
Nd$_{2}$Fe$_{14}$B, revealed by quantitative Kerr and Faraday 
microscopy, propagate into the bulk and exhibit structural 
features with dimensions down to 6 nm, the domain wall thickness. 
The observed fractal nature of the domain structures provide an 
explanation for the anomalous increase in the bulk magnetization 
of Nd$_{2}$Fe$_{14}$B below the spin-reorientation transition. 
These measurements open up a rich playground for studies of 
fractal structures in highly anisotropic, magnetic systems.
\end{abstract}


\maketitle{}

The industrial strength ferromagnet, Nd$_{2}$Fe$_{14}$B has 
become a prototypical system for the study of magnetic domain 
structures. Below the Curie temperature, 
$T_c$~=~565~K\cite{Buschow86,Herbst91}, the Nd and Fe moments 
order ferromagnetically. The crystal-electric field produces a 
strong magnetic anisotropy with the easy axis along the 
tetragonal $\mathbf{c}$ direction. Below the spin-reorientation 
temperature, $T_{\mathrm{SR}}$~=~135~K, the magnetic structure 
(and easy axis direction) changes via a second order 
transition\cite{Pastushenkov02} and becomes 
cone-like\cite{Inaba88}, in which the moments are canted away 
from the $\mathbf{c}$ direction by an angle that increases from 
0~deg at $T_{\mathrm{SR}}$ to 28~deg at 
4~K\cite{Tokuhara85,Pastushenkov97}. The moments lie in one of 
the four symmetry-equivalent \{1~1~0\} planes in agreement with 
calculations of crystal-electric field effects\cite{Yamada88}.  A 
multitude of techniques have been used to image magnetic domains 
at exposed surfaces of Nd$_{2}$Fe$_{14}$B, such as Bitter 
decoration\cite{Corner88}, Kerr 
microscopy\cite{Pastushenkov02,Pastushenkov97,Szymczak87,Folks94,Hubert00}, 
Lorentz\cite{Lemke97,Zhu98,Shinba05} and 
holographic\cite{Zhu98,Park04} transmission electron microscopy, 
scanning electron microscopy\cite{Zhu98,Wang98}, and magnetic 
force 
microscopy\cite{Lemke97,Gruetter88,Neu04,Szmaja06,AlKhafaji98}. 
For imaged surfaces perpendicular to the $\mathbf{c}$ direction, 
domains have been observed with dimensions between 2-5~$\mu$m at 
4~K\cite{Pastushenkov97} and about 0.1-0.6~$\mu$m at room 
temperature\cite{Pastushenkov97,Folks94,Lemke97}, respectively. 
These values are very close to the single domain size of about 
0.2-0.4~$\mu$m determined by magnetization measurements on 
polycrystalline samples\cite{Buschow86,Groenefeld90}.

These imaging techniques, however, only probe the magnetic fields 
at the surface and/or are limited in terms of the size of fine 
details that can be resolved.  In the case of transmission 
electron microscopy, for example, samples thinner than 1~$\mu$m 
are required and the domain structure can be very different from 
that of a bulk specimen.  Thus, important questions regarding the 
size, distribution and morphology of magnetic domains in the bulk 
remain open.  Here, we address these issues by correlating the 
results of magnetization measurements, magnetic domain imaging, 
and small-angle neutron scattering experiments on solution grown 
Nd$_{2}$Fe$_{14}´^{11}$B single 
crystals\cite{Canfield92,Canfield01} that manifest smooth, 
mirrored surfaces and can have volumes as large as 1~cm$^{3}$.

Single crystalline Nd$_{2}$Fe$_{14}´^{11}$B shows a strong
increase in magnetization along the $\mathbf{c}$ direction below
$T_{\mathrm{SR}}$~=~135~K, similar to what has been reported
previously\cite{Herbst91,Tokuhara85}.  As shown in Fig. 1, this
behavior is correlated with a dramatic change of the magnetic
domain patterns above and below $T_{\mathrm{SR}}$ as measured by
the magneto-optical Kerr effect.  In both the high and the low
temperature states, the domains are arranged in chains in the
($\mathbf{ab}$) plane.  However, the fine domains observed above
$T_{\mathrm{SR}}$ evolve to much larger domains below
$T_{\mathrm{SR}}$.  Above $T_{\mathrm{SR}}$, the domains exhibit
more isotropic star-like boundaries whereas, below
$T_{\mathrm{SR}}$, the domains are nearly rectangular in shape.
The Faraday microscopy patterns, shown in Fig. 2, provide a
spatially resolved measurement of the local magnetization along
the $\mathbf{c}$ direction.  Above $T_{\mathrm{SR}}$, higher 
values for the magnetic induction perpendicular to the surface in 
the inner parts of domains are observed, indicated by the red 
regions. This is consistent with the uniaxial alignment of 
moments along the $\mathbf{c}$ direction, in contrast to the 
canted moments below $T_{\mathrm{SR}}$ for which only the moment 
projection along the $\mathbf{c}$ direction contributes.  This 
result appears to be in contradiction to the decreased bulk 
magnetization measured along the $\mathbf{c}$ direction above 
$T_{\mathrm{SR}}$.  We show below that this arises from the 
details of the domain structure within the bulk of the 
Nd$_{2}$Fe$_{14}´^{11}$B single crystal.

\begin{figure}
\includegraphics[width=0.8\linewidth]{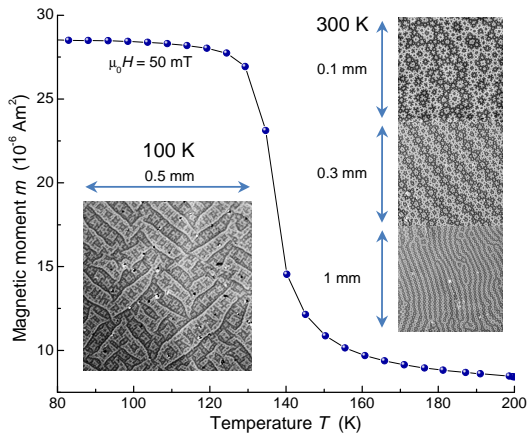}
\caption{\label{fig:fig1} Temperature dependence of the 
magnetization and the magnetic domain patterns in the 
Nd$_{2}$Fe$_{14}´^{11}$B single crystal.  The magnetization was 
measured at $\mu_0 H$~=~50~mT applied along the $\mathbf{c}$ 
axis. The magnetic domain patterns were imaged exploiting the 
magneto-optical polar Kerr effect at a surface perpendicular to 
the $\mathbf{c}$ direction and with the $\mathbf{a}$ direction 
vertical.}
\end{figure}

\begin{figure}
\includegraphics[width=0.75\linewidth]{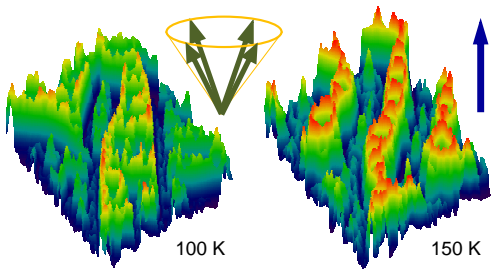}
\caption{\label{fig:fig2} Faraday images of the 
Nd$_{2}$Fe$_{14}´^{11}$B single crystal.  In the transparent 
ferromagnetic Bi-doped Fe-garnet film placed on the surface of 
the sample, the distribution of the magnetic moments mimics and 
amplifies the magnetic induction on the sample surface.  The 
height is proportional to the magnitude of the magnetic induction 
in $\mathbf{c}$ direction perpendicular to the surface.  The spin 
structure is shown schematically.}
\end{figure}

Figures 1 and 2 demonstrate clearly that domain features on many
length scales can be detected at the surface of the sample.  But
do these features extend into the bulk?  With a sufficiently 
large penetration depth and sensitivity to magnetic moments and 
magnetostrictive effects, neutrons are well suited to study this 
material using small-angle neutron scattering (SANS) methods.  
SANS measurements made with the incident neutron beam parallel to 
the $\mathbf{c}$ axis of the sample yield a two-dimensional 
Fourier transform of magnetic and magnetostrictive correlations 
within the ($\mathbf{ab}$) plane of the sample, averaged over the 
thickness of the sample.  Using well-established 
techniques\cite{Lindner92}, the average dimension and 
conformation of domains can be probed, and characteristic length 
scales of the magnetic induction variation can be extracted from 
the measurements.  The minimum scattering vector of 
$\sim$0.5$\cdot10^{-3}$~\AA$^{-1}$, that can be realized by 
conventional SANS instruments, limits the dimension of detectable 
features to length scales smaller than $\sim$1~$\mu$m, which is 
just at the limit of optical techniques.

SANS experiments were performed on the D11 
instrument\cite{Lindner92}, at the ILL, on a plate-like 
Nd$_{2}$Fe$_{14}´^{11}$B sample with its $\mathbf{c}$ direction 
oriented parallel to the incident neutrons.  Fig. 3 shows the raw 
SANS area-detector images both above and below $T_{\mathrm{SR}}$. 
Even a cursory comparison of the two images reveals significant 
differences.  First, the pattern below $T_{\mathrm{SR}}$ exhibits 
significantly more anisotropy than that above $T_{\mathrm{SR}}$ 
which is consistent with the rectangular habit of magnetic 
domains for $T<T_{\mathrm{SR}}$ shown in the left panel of Fig. 
1.  By comparing the intensity scales in Fig. 3, it can be seen 
that the pattern at $T$~=~200~K > $T_{\mathrm{SR}}$, exhibits 
significantly enhanced scattering. This indicates a strong 
decrease in the average domain size above $T_{\mathrm{SR}}$.  
Perhaps most importantly, since SANS probes the distribution of 
magnetic domains throughout the sample, these domain 
distributions are characteristic of the bulk of the 300~$\mu$m 
thick sample, not only the exposed surface.

\begin{figure}
\includegraphics[width=0.95\linewidth]{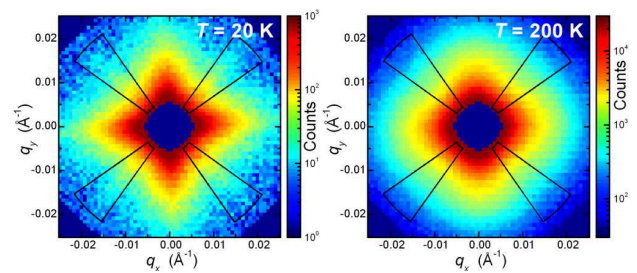}
\caption{\label{fig:fig3} Raw SANS area-detector images of the
Nd$_{2}$Fe$_{14}´^{11}$B single crystal at temperatures of 20~K
and 200~K.  The colors and lines in the contour map represent the 
neutron counts on a logarithmic scale.  The black lines mark the 
20~deg wide sectors used for the averaging to obtain the data 
shown in Fig. 4.}
\end{figure}

SANS patterns have been measured over a wide range of scattering
vectors and the scattered intensity was determined as function of
the magnitude of the scattering vector.  The results are 
displayed in Fig. 4.  Information regarding the formation of the 
magnetic domains can be obtained by fitting regions of the SANS 
data to a power-law behavior of the form 
$I~=~I_{0}~\cdot~q^{\alpha}$.  For details concerning the 
interpretation of power laws in SANS measurements, we refer the 
reader to several excellent reviews in Refs. 
\cite{Schmidt95,Mildner86,Schmidt89,Martin87}.  Although we 
follow their treatments here, we note that much of the previous 
work in this area is based upon systems consisting of well 
defined ``objects'' with a finite homogeneous scattering strength 
embedded in a ``matrix'' with a different scattering strength.  
The situation for scattering from magnetic domains is somewhat 
different since the ``objects'' are the interior regions of 
domains with an identical and constant absolute value of the 
scattering strength that can be positive or negative depending on 
the direction of the local magnetization.  The ``surfaces'' of 
these ``objects'' are the domain walls, with their own magnetic 
structure that varies across the domain wall width.  Together, 
the domains and domain walls fill the complete volume of the 
sample, so no ``matrix'' in the sense used above exists.

\begin{figure}
\includegraphics[width=0.65\linewidth]{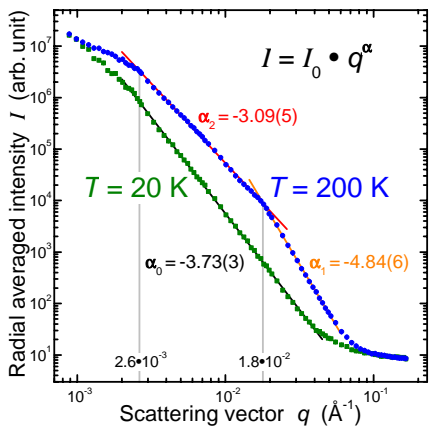}
\caption{\label{fig:fig4} Scattering vector dependence of the 
SANS intensity of the Nd$_{2}$Fe$_{14}´^{11}$B single crystal at 
temperatures of 20~K and 200~K.  These data were determined by 
averaging over all four sectors marked in Fig. 3, and combining 
patterns measured at different sample-to-detector distances and 
different energies of the incident neutrons.  The intensities 
extracted from sectors rotated by 45~deg to the sectors outlined 
in Fig. 3 are, on average, a factor of two larger, but show a 
similar dependence on the scattering vector.}
\end{figure}

At $T$~=~20~K, the neutron scattering intensity shows a power-law 
dependence over a wide scattering-vector range with a single 
exponent $\alpha_{0}$ from $\sim$2$\cdot10^{-3}$~\AA$^{-1}$ up to 
$\sim$3$\cdot10^{-2}$~\AA$^{-1}$.  The value of 
$\alpha_{0}=-3.73(3)$ is not far from $\alpha$~=~-4 which would 
be the expected exponent for scattering from a ``smooth 
surface'', the so-called Porod 
scattering\cite{Schmidt95,Mildner86,Schmidt89,Martin87}.  A 
wrinkle-free, sharp domain wall is the corresponding picture in a 
magnetic system.  Besides scattering from magnetic domains, 
neutrons can be refracted at domain walls due to the difference 
in the refraction index on both sides of the domain wall.  This 
spreading of the incident beam can cause intensity for finite 
scattering vectors in SANS experiments and is strongly wavelength 
dependent\cite{Schaerpf78, Hoffmann07}.  Contributions to the 
observed intensity can not completely excluded but measurements 
for scattering vectors between $\sim$2$\cdot10^{-3}$~\AA$^{-1}$ 
and  $\sim$1$\cdot10^{-2}$~\AA$^{-1}$ using neutrons with  
wavelengths between 4$~\AA$ and 17$~\AA$ yield only marginal 
differences in the normalized intensity and deviations in 
$\alpha$ of smaller than 0.2. 

At $T$~=~200~K, we see dramatic changes in the intensity and 
shape of the scattering curves.  The number of neutrons scattered 
in the range between $\sim$1$\cdot10^{-3}$~\AA$^{-1}$ up to 
$\sim$1$\cdot10^{-1}$~\AA$^{-1}$ is, on average, a factor of six 
higher at $T$~=~200~K than at $T$~=~20~K.  A distinctive kink is 
obvious at $\sim$1.8$\cdot10^{-2}$~\AA$^{-1}$ in the curve for 
$T$~=~200~K in contrast to the smooth shape for $T$~=~20~K in the 
double-logarithmic plot in Fig. 4.  Neutron refraction from an 
increased density of domain walls can yield an intensity increase 
but would not cause such a distinct kink.  The range of 
scattering vectors corresponds to fine-scale features of domains 
from $\sim$0.1~$\mu$m down to $\sim$0.001~$\mu$m, well below the 
resolution limit of magneto-optical techniques, but close to the 
resolution limit of magnetic force 
microscopy\cite{Szmaja06,AlKhafaji98}.  These features are also 
significantly smaller than the domain sizes described by the 
branching phenomena in magnetic 
domains\cite{Szymczak87,Hubert00}: for a 300~$\mu$m thick 
Nd$_{2}$Fe$_{14}$B single crystal, the bulk domains, at half 
sample thickness, are expected to be $\sim$10~$\mu$m 
wide\cite{Szymczak87,Hubert00,Kronmueller03}, whereas a width of 
$\sim$0.6~$\mu$m was determined for the ``spike'' domains in 
near-surface regions\cite{Szymczak87,Hubert00,Kronmueller03}.  In 
addition to these coarse domain structures, a complicated system 
of fine-scale surface domains with the topology of curved stripes 
with widths of 0.02-0.25~$\mu$m has also been observed by 
magnetic force microscopy\cite{Szmaja06,AlKhafaji98}.  Domain 
features on this length scale are consistent with the observed 
signal in the reported SANS measurement.

For a more detailed discussion of the SANS results at 
$T$~=~200~K, it is convenient to consider two distinct $q$ ranges 
denoted by the vertical lines in Fig. 4. For scattering vectors 
$q>q_{\mathrm{min,1}}=1.8\cdot10^{-2}~\AA^{-1}$, the scattering 
describes the properties of the domain 
``surface''\cite{Schmidt95,Mildner86}.  The observed value of 
$\alpha_{1}$~=~-4.84(6) in the power law is significantly 
different from $\alpha$~=~-4 for a sharp domain wall, with an 
abrupt change in the direction of the 
magnetization\cite{Schmidt95,Mildner86,Schmidt89,Martin87}.  It 
falls in the class, $\alpha=-(4+2\beta)$ with $0<\beta<1$ 
characteristic of a ``smooth pore boundary''\cite{Schmidt95}, in 
which the scattering strength changes continuously over some 
distance by a power law with the power $\beta$.  The value of 
$\beta$~=~0.42(3) corresponds roughly to a square-root dependence 
of the scattering strength in the ``boundary''\cite{Schmidt95}, 
close to the sinusoidal dependence expected for Bloch or N\'{e}el 
domain walls.  The lower limit 
$q_{\mathrm{min,1}}=1.8\cdot10^{-2}~\AA^{-1}$ for this observed 
power-law dependence is related to the width of the domain 
wall\cite{Mildner86} and yields a value of order 
$1/q_{\mathrm{min,1}}$~=~6~nm which is in excellent agreement 
with values of 4-9~nm determined by other 
methods\cite{Buschow86,Corner88,Zhu98,Gruetter88}.  The increased 
scattering evident at $T$~=~200~K in this region, indicates that 
a significantly larger volume fraction of the sample is found in 
domain walls above $T_{\mathrm{SR}}$ and provides an explanation 
for the reduced magnetization above $T_{\mathrm{SR}}$ since the 
domain walls should not, or only marginally, contribute to the 
magnetization of the sample.  With a strong increase in the 
volume fraction of domain walls in the bulk of the sample, the 
total magnetization measured parallel to the $\mathbf{c}$ axis is 
reduced above $T_{\mathrm{SR}}$ even though the moments are more 
closely aligned along the $\mathbf{c}$ axis.  

In the intermediate scattering vector range, 
$2.6\cdot10^{-3}~\AA^{-1}=q_{\mathrm{min,2}}<q<1.8\cdot10^{-2}~\AA^{-1}=q_{\mathrm{max,2}}$, 
a power $\alpha_{2}$~=~-3.09(5) was obtained from the SANS data 
measured at $T$~=~200~K.  The observation of a power law over a 
scattering vector range of one decade with an exponent $\alpha$ 
close to -3 characterizes the scattering ``objects'' as 
fractals\cite{Schmidt95,Mildner86,Schmidt89,Martin87,Fractal}. 
Indeed, the fractal nature may be expected for magnetic domains 
as we move from the two-dimensional fine surface pattern observed 
by Faraday imaging\cite{Han02} toward the coarse 
three-dimensional bulk domain structures via the branching 
process\cite{Hubert00}. However, the present SANS data 
characterize the fractal nature of magnetic domains on a 
submicron length scale in the bulk.  Interestingly, an exponent 
of $\alpha\sim-3$ in the power law for the scattering data marks 
the boundary between two types of 
fractals\cite{Schmidt95,Mildner86,Schmidt89,Martin87}, surface 
fractals with $-4<\alpha<-3$ and volume fractals with 
$-3<\alpha<-1$.  A surface fractal with $\alpha$ close to -3 can 
be described as a self-affine ``surface'', which is so strongly 
wrinkled, that it almost fills the three dimensional space.  The 
corresponding picture for a volume fractal with $\alpha\sim-3$ is 
that, starting from a point in a self-affine ``object'', one can 
find a minimum of one other point in a given maximum distance 
$r$, which is not within the same ``object''.  The range for $r$, 
over which this self-affinity can be found, corresponds to the 
scattering vector range in which the specific power law is 
observed.  Regardless of the starting point, both descriptions 
yield the same picture.  The strong neutron scattering intensity 
for the intermediate scattering vector range is related to 
magnetic domains in the bulk with self-affinity on a length scale 
from $1/q_{\mathrm{max,2}}$~=~6~nm to roughly 
$1/q_{\mathrm{min,2}}$~=~40~nm.

In summary, for the case of strong uniaxial anisotropy in 
Nd$_{2}$Fe$_{14}$B, we see that very fine magnetic domains occur 
at the surface and in the bulk with a self-affine geometry.  
Below the spin reorientation transition, the domain structure is 
much coarser, less wrinkled and the effective volume of domain 
walls is smaller.  This leads to an apparent increase of the 
measured magnetization along the $\mathbf{c}$ axis even though 
the local magnetization decreases.  Small-angle neutron 
scattering is not only a valuable tool to characterize the 
conformation of magnetic domains and their domain walls, it also 
provides information regarding the topology of magnetic domain 
arrangements.  Finally, we note that the magnetic properties and 
the corresponding magnetic domain patterns can be tuned by 
varying external parameters (e. g. temperature, pressure or 
applied magnetic field) as well as internal parameters (e. g. 
change of the magneto-crystalline anisotropy by substitution of 
different rare earths).  These modifications can be isotropic or 
anisotropic providing a rich and almost unique opportunity for 
experimental studies of (magnetic) fractals.

The authors are indebted to Y. Janssen and S. Nandi for
stimulating discussions.  Work at the Ames Laboratory was
supported by the Department of Energy, Basic Energy Sciences,
under Contract No. DE AC02-07CH11358.  A. Kreyssig acknowledges
the support by Deutsche Forschungsgemeinschaft through SFB 463.
R. Prozorov acknowledges support by the NSF Grant No. DMR
05-53285 and by the Alfred P. Sloan foundation.


\begin{thebibliography}{}
\bibitem[()]{Fractal}A fractal object remains ``similar'' to itself after an isotropic change of scale or dilation.  For ``self-affine'' objects, this dilation can be anisotropic with different scaling factors for different directions$^{26}$.
\end{thebibliography}
\end{document}